\documentstyle[aps,epsfig,prb,floats]{revtex}
\newcommand{\rmu}{\mathrm{\mu}}
\newcommand{\etal}{{\it et al.}}

\begin{document}
\draft

\twocolumn[\hsize\textwidth\columnwidth\hsize\csname @twocolumnfalse\endcsname

\title{Electronic transport in films of colloidal CdSe nanocrystals}
\author{Nicole Y. Morgan$^1$, C.A. Leatherdale$^2$,  M.
Drndic$^1$, Mirna Vitasovic$^2$, Marc A. Kastner$^1$, Moungi Bawendi$^2$,}
\address{$^1$Dept. of Physics, $^2$Dept. of Chemistry, M.I.T., 77
Massachusetts Ave., Cambridge MA 02139}
\date{\today}
\maketitle

\begin{abstract}

We present results for electronic transport measurements on large
three-dimensional arrays of CdSe nanocrystals.  In response to a step
in the applied voltage, we observe a power-law decay of the current
over five orders of magnitude in time.  Furthermore, we observe no
steady-state dark current for fields up to $10^6$ V/cm and times as
long as $2\times10^4$ seconds.  Although the power-law form of the
decay is quite general, there are quantitative variations with
temperature, applied field, sample history, and the material
parameters of the array.  Despite evidence that the charge injected
into the film during the measurement causes the decay of current, we
find field-scaling of the current at all times.  The observation of
extremely long-lived current transients suggests the importance of
long-range Coulomb interactions between charges on different
nanocrystals.

\end{abstract}
]
\subsection*{Introduction}

The techniques of colloidal chemistry make it possible to create vast
numbers of nearly identical semiconductor nanocrystals
\cite{baw.ncrev,ncrev}.  Measurements of the tunneling of electrons
onto individual nanocrystals have demonstrated that the latter behave
qualitatively like lithographically defined quantum dots
\cite{cdseset}; such structures are best described as artificial atoms
because their energy and charge are quantized
\cite{ashoori96,kastner.phystoday}. Although it is difficult to
construct large arrays of lithographically patterned artificial atoms,
colloidal nanocrystals assemble themselves into such arrays quite
naturally, creating an entirely new class of solids composed of
artificial atoms.  Furthermore, by adjusting the chemical process, it
is possible to tune physically relevant parameters, such as site
energies and nearest-neighbor coupling.  We report here studies of
electron transport in such a Colloidal Artificial Solid (CAS).

Our CAS is composed of CdSe nanocrystals $\sim5$ nm in diameter, each
capped with organic molecules $\sim1$ nm long.  Because of the small
size of the nanocrystals the Coulomb interaction between two electrons
on the same nanocrystal or adjacent nanocrystals is $\sim0.1$ eV
\cite{hodes}, larger than $k_BT$ at room temperature. It is likely,
therefore, that the Coulomb energy is larger than the other important
energies in the problem, in particular the bandwidth resulting from
inter-nanocrystal tunneling or disorder.  Thus, these CASs provide an
interesting new system in which the motion of electrons is expected to
be highly correlated \cite{levitov}.  

Although the optical properties of these nanocrystal systems have been
studied extensively, as yet there has been little work on electronic
transport in these arrays.  In this work, we present measurements made
on a lateral, gated device, shown in Fig.\ \ref{figsetup}; this
geometry has two primary advantages over the simpler sandwich
structure used by Ginger and Greenham \cite{ginger}.  First, because
the nanocrystal films are deposited last, there is no possibility of
damaging the films during device fabrication.  Second, the gate
electrode provides the possibility of controlling the charge density
in the CAS.

We find that films of CdSe CASs are extraordinarily resistive, with
resistivity greater than $\sim10^{14}$ ohms-cm at temperatures below
200 K.  Although there is no detectable steady-state current, we
observe current transients after the application of a voltage step;
the current decays as a power law in time out to at least $10^4$ s.
When the film is located near a metal gate, we can directly observe
the buildup of charge in the sample, which, at low temperatures, then
remains in the CAS even after the voltage is removed.  Although this
suggests that the decay of the current is associated with the
charging, the integral of the current transient is orders of magnitude
larger than the charge stored.  Thus, we conclude that the current
arises from charge which moves completely through the film, but that
the sample becomes more resistive with time as more charge is
injected.  We suggest that these effects are related to the strong
Coulomb correlations between electron occupancy of the nanocrystals.
In this paper, we present transport measurements on this system as a
function of device geometry, applied field, sample length, and
temperature, as well as some preliminary data on variations of the
transients with the parameters of the chemical synthesis.

\subsection*{Experimental details}

The synthesis of the CdSe nanocrystals and the deposition of the
close-packed films has been described in detail previously
\cite{murray,murray.sci}.  By using the organo-metallic synthesis of
Murray, \etal\ followed by three steps of size-selective
precipitation, a size distribution with variance $\sigma = 5-8\% $ is
typically obtained.  For most of the measurements in this paper, the
nanocrystals are covered with a monolayer of tri-octylphosphine oxide
(TOPO), as a byproduct of the synthesis.  These molecules serve to
passivate the surface states of the nanocrystals, and also act as
nearest-neighbor spacers when the nanocrystal films are formed.  For
some samples, after the synthesis the capping molecules have been
exchanged from TOPO to tri-butylphosphine oxide (TBPO), which gives a
smaller nearest-neighbor separation; the procedure for the exchange
has been described elsewhere \cite{calphoto}.  For the work discussed
here, the nanocrystals typically have a core diameter of approximately
4.5 nm; those capped with TOPO have a 1.1 nm nearest-neighbor
separation, and those with TBPO have a 0.7 nm nearest-neighbor
separation \cite{murray.sci}.  The films are formed by drop-casting a
solution of nanocrystals in 9:1 hexane/octane, which is allowed to dry
for a minimum of two hours in an inert atmosphere before exposure to
vacuum.  To a first approximation, the thickness of the films is
controlled by the concentration of the solution; for the work reported
here, the thickness of the films ranges between 20 and 180 layers
($100-900$ nm), as measured with a profilometer near the measurement
electrodes.  Direct imaging of the nanocrystal packing on the
measurement substrates has proven difficult, but transmission electron
microscope images of similarly prepared films suggest that the
ordering within the CAS is polycrystalline, with the typical grain
size approximately 10 nanocrystals across \cite{drndic.tbp}.

\begin{figure}
\epsfig{file=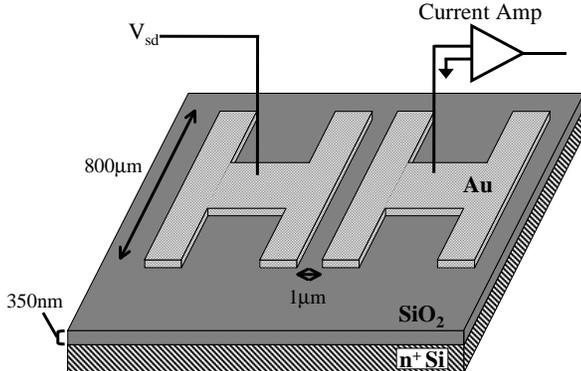, width=0.43\textwidth}
\caption{A schematic of the most commonly used substrate and electrode
structures.  The base material is a standard commercial (100) silicon
wafer, degenerately doped with arsenic.  The 350 nm thick oxide is
thermally grown, and the electrodes (200 \AA\ Ti, 2000 \AA\ Au) are
patterned with a single optical lithography step and a lift-off
technique.  When the nanocrystal film is deposited, it covers the
entire surface, including the electrodes. The associated measurement
circuit is described in the text. }
\label{figsetup}
\end{figure}

The primary type of device structure used for this work is shown in
Figure \ref{figsetup}.  These structures have been produced in large
quantities so that many films, some with different core sizes and
different capping molecules, can be deposited on identical devices.
The starting material is a (100) silicon wafer degenerately doped with
arsenic (room temperature sheet resistance 1--5 m$\Omega$), which is
used as the back gate for the device.  High-quality thermal oxides of
350--500 nm are grown on top of these substrates, and then 200 nm
Ti-Au electrodes are patterned and deposited using optical lithography
and a lift-off procedure.  We have also made measurements on films
deposited on 0.5 mm quartz wafers, for which the gold electrode
pattern is identical, but for which the electrodes are only 110 nm
thick.  In both cases, although only one set of electrodes is depicted
in Fig.\ \ref{figsetup}, there are many more electrodes, some with
different gap sizes, on each wafer.  The full size of the substrate is
typically $6\times6$ mm, and the size affects the spreading of the
nanocrystal solution during film deposition. In addition, for some
measurements we have used ring-shaped electrodes (not shown)
with a similar gap size for a few experiments, for which the results
are consistent.

The fabrication of the substrates is completed before the nanocrystals
are deposited and then dried overnight in an inert atmosphere;
afterwards, the nanocrystal film covers the entire substrate,
including the electrodes.  Depositing the nanocrystals as the final
step in sample preparation is important.  These films are not
sufficiently robust to permit the use of standard wafer processing
techniques after deposition.  In particular, subsequent measurements
have shown that even moderate heating, to $100\deg$ C in vacuum, can
significantly affect the transport and optical properties of the films
\cite{drndic.tbp}.  The electrode geometry and the measurements are chosen
to give a well-defined active area for the device, as will be
discussed below.  For the most common samples, 4.5 nm core-diameter,
TOPO-capped nanocrystals, and the electrode configuration of Fig.\
\ref{figsetup} there are approximately 200 dots in series and about
160,000 dots in parallel for one layer of dots.  Room-temperature
optical measurements of nanocrystal films co-deposited on glass slides
are used to ensure sample quality.

The electrical measurements are made in an Oxford Variox cryostat,
with the sample in vacuum, typically at liquid nitrogen temperature.
The current is measured with a Keithley 427 current amplifier;
depending on the voltage and time ranges of interest, one of several
voltage sources, including a Yokogawa 7611, a HP 3245, and a Kepco
BOP-500V, is used.  In a typical measurement, a voltage step is
applied at one of the electrodes, and the current is measured at the
other electrode as a function of time, with the back gate grounded.
Throughout this paper we will refer to the electrode at which the
voltage step is applied as the source, and to the current measurement
electrode as the drain, as indicated in Fig.\ \ref{figsetup}.  For the
gated silicon substrates, this measurement is asymmetric in voltage
when the gate is grounded; because the drain electrode is connected to
the input of the current amplifier, which is a virtual ground, the
field at the source is much larger than that at the drain.  As a
result, charge is only injected into the film from the source
electrode for the gated substrates.  Furthermore, although charge
spreads into the film in all directions from the source, the
measurement at the drain electrode is sensitive only to charge moving
between the electrodes, as more distant charge will be screened by the
gate \cite{fn0}. Finally, this measurement is not directly sensitive
to leakage currents between the source and the gate, and there is no
field to generate leakage current between the drain electrode and the
gate.

\subsection*{Results}

\begin{figure}
\epsfig{file=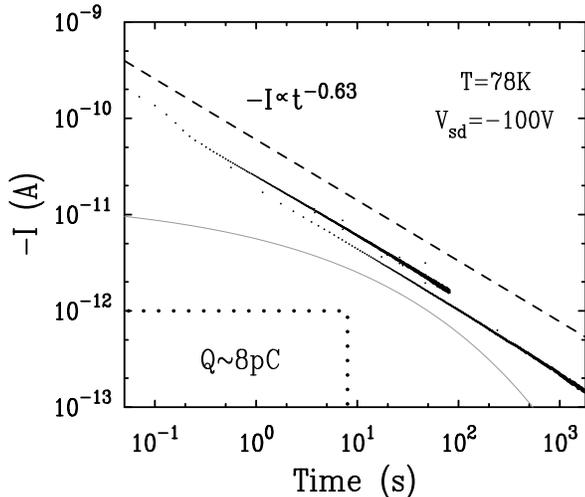, width=0.43\textwidth}
\caption {A typical current transient, measured for a CAS with 45 \AA\
core-diameter, TOPO-capped nanocrystals on a substrate with a 350 nm
gate oxide, at 78K.  With the gate grounded, the source voltage is
stepped to $-100$ V, and the current at the drain electrode is
measured as a function of time.  Two measurements, with different
circuit response times, are shown; the difference in the amplitudes of
the two transients reflects sample history effects. The dashed line
represents a power-law decay of the current, with an exponent of
$-0.63$.  The rectangle outlined by a dotted line in the lower left
corner represents an upper bound for the amount of charge expected
from capacitive charging of the quantum dot film in the gap between
the measurement electrodes.  The deviation from power-law behavior at
short times varies with the series resistance and is therefore
ascribed to charging of geometrical capacitances.}
\label{figlm1}
\end{figure}

As mentioned above, the CAS films are highly resistive.  For a film of
4.5 nm diameter CdSe nanocrystals we measure no steady-state current
for applied fields of up to $10^6$ V/cm, which corresponds to an
average nearest-neighbor potential difference of 500 mV. With a
current detection limit of 5 fA, and the device geometry used here,
this gives a lower bound for the resistivity of of $\sim10^{14}
\Omega$-cm; no steady state current has been observed in any of the
dozens of samples we have studied.  However, the time dependence of
the current in response to an applied source-voltage step is unusual.
Figure \ref{figlm1} shows a plot of the typical current measured at
the drain electrode as a function of time; unless otherwise noted, the
gate is grounded in all of the transient measurements. At $t=0$, the
voltage at the source electrode has been stepped from zero to -100 V,
after which the resulting current decays with a power-law form.  In
other measurements, the power-law decay of the current has been
observed to persist overnight, to approximately $2\times 10^4$
seconds.  Together with the data are plotted two fits, displaced
vertically for clarity.  Above the data is a power law, $I = I_1
t^{-\alpha}$, with $\alpha = -0.63$, and below the data is a stretched
exponential $I = I_0 \exp{-(t/\tau)^{\beta}}$, with the characteristic
time $\tau $ set at 1 ms.  Clearly, the data are better described by a 
power law form. \cite{fn1}

For the various nanocrystal samples measured, the decay exponent,
$\alpha$, ranges from -0.1 to -1; the parameters upon which $\alpha$
depends are discussed below.  Thus, the charge obtained from
integrating the transient current diverges in time.  By itself, this
divergence suggests that the current is not just the displacement
current that arises from the polarization or to the buildup of charge
in the CAS.  In fact, the value of the integrated current over a
typical measurement time of $10^3$ seconds is several orders of
magnitude greater than the estimate for charge stored in the film.  To
illustrate this, in the lower left of Figure \ref{figlm1}, we have
drawn a dotted rectangle, which represents the amount of charge that
would reside in the film, if the entire active area between source and
drain were charged to the full potential of the applied voltage step,
$V_{sd}$ with respect to the gate electrode. \cite{fn2} Here, this
estimated charge is approximately 8 pC, whereas for the transient
shown in Fig.\ \ref{figlm1}, the integrated current exceeds this value
by more than two orders of magnitude.  Were this measurement not
truncated the disparity would be greater still.  We conclude that the
current arises from charge which travels through the sample, and that
the current decays because the film grows more resistive with time.

Nonetheless, we believe that the increase of the resistance over time
is related to the buildup of charge in the CAS.  The injection of
charge into the film can be directly measured by applying voltage to
the gate electrode while measuring the current at one of the top
electrodes; for this type of measurement the source and drain are both
held at ground potential and are therefore equivalent.  As noted
above, there are many other electrodes on the top of the oxide in
contact with the CAS film, and these are all at the same potential as
the source and drain.  Although charge spreads into the film from all
of these electrodes, typically the current is measured only at one of
them; furthermore, this current arises only from the charging of the
nanocrystal film.

\begin{figure}
\epsfig{file=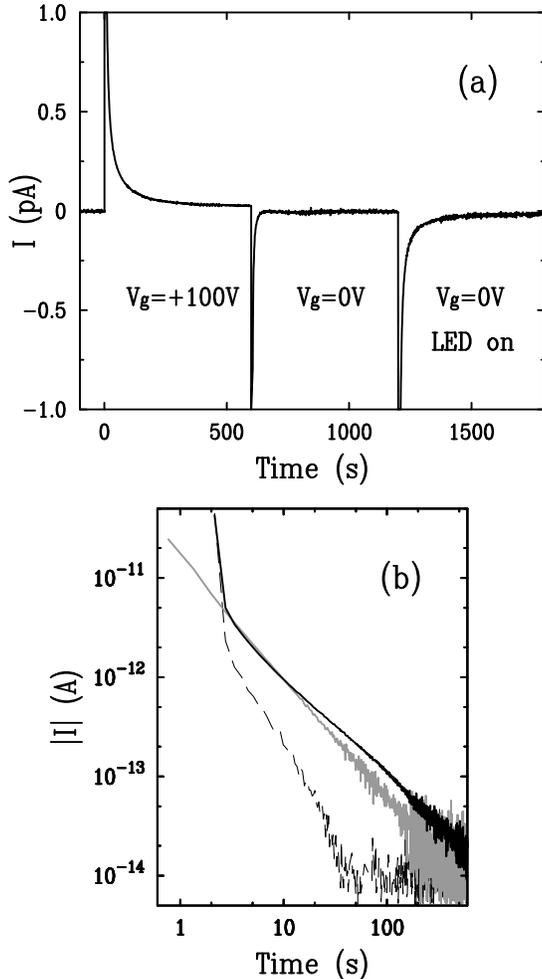, width=0.4\textwidth}
\caption{(a)Current transients measured at one of the top electrodes
for a series of ten-minute gate voltage steps at 78K.  The measurement
electrode is grounded through the current amplifier; all other top
electrodes are directly grounded throughout.  Integrating the current,
including the fast capacitive component, gives 85 pC for the first
step, in which $V_g$ is stepped to $+100$V; $-34$ pC for the second
step, in which $V_g$ is stepped back to 0 V; and $-46$ pC for the
third step, in which $V_g$ is held at zero and the sample exposed to
band-gap light.  The 34 pC in the second step is consistent with
estimates of the charge on the geometrical capacitance. (b) The
absolute value of the same three current transients, plotted on a
logarithmic scale.  The first transient ($V_g\rightarrow100$ V) is
shown in black, and clearly shows a fast capacitive component at short
times and an approximate power-law decay at long times.  The second
transient ($V_g\rightarrow0$ V) is shown as a dashed line, and shows
only the fast capacitive component, followed by a very small component
attributed to polarization of the substrate.  Transients measured for
$V_g\rightarrow-100$ V are identical to this second transient, i.e.,
they show no evidence of charge injection.  The third transient (LED
on) is shown in gray, and decays as a power-law with an exponent of
$-1.2$.  Note the absence of the capacitive component, as well as the
similarity between the time dependence of the discharging and that of
the charging in the first transient.}
\label{figgatecharge}
\end{figure}

The response of the sample to the gate voltage is shown in Figure
\ref{figgatecharge}, for three consecutive ten minute steps.  In the
first step, the gate voltage is stepped to +100 V, and negative charge
flows slowly into the film from the top electrodes; after ten minutes
the current is still nonzero.  In addition to this slow charging, at
short times there is also a contribution from the displacement
current, which corresponds to the charging of the capacitance between
the measuring electrode and the gate.  In the second step, when the
gate voltage is stepped back to zero, only this displacement current
is observed.  Any current that might indicate a discharge of the
nanocrystal film is below the noise floor, which is approximately 0.01
pA for this measurement.  After being held with $V_g=0$ V for ten
minutes, in the third step the sample is illuminated with band gap
light from a light-emitting diode (LED) near the sample in the
cryostat.  The wavelength of the LED has been chosen to match the
absorption peak of the CAS. This illumination causes the charge to be
released from the film.  The integral of the total current for each
time period is indicated in the figure caption.  The total charge
integrated over the entire process is very small, which indicates that
most of the charge which was injected in the +100 V step is removed by
exposure to the LED. \cite{fn3}

Figure \ref{figgatecharge}b shows the same data for the charging and
discharging of the CAS on a logarithmic scale.  Interestingly, the
time decay of the discharge under illumination is very similar to that
of the charging; in both, the current has an approximate power law
form, although the decay here is steeper than for the source-drain
measurements.  Since the decay of the current follows approximately
$t^{-1}$, the charge spreads only logarithmically in time.  From the
perimeter of the H-shaped electrodes, approximately 4.4 mm, and the
350 nm thickness of the gate oxide, we calculate the surface charge
density at which the nanocrystal film is charged to 100 V. 
Using this calculated density and the value of the stored charge, 46
pC, measured from the light-induced discharge, we estimate that the
injected charge has spread out roughly $1.1 \rmu$m from the electrode
during the ten-minute charging step.

There is also a clear difference between the current during the first
step, with $V_g=+100$ V, and that in the second step, for which the
voltage is stepped back to zero.  In the latter transient, the
expected capacitive transient is seen at short times ($\leq3$ s).  At
longer times there is a low-level current which decays relatively
rapidly.  This background transient, unlike the charging of the film,
is linear in voltage; we believe that it results from polarization of
the oxide.  Like the measurements of Fig.\ \ref{figlm1}, These gate
charging measurements exhibit a strong asymmetry with the sign of
the applied voltage.  When the gate voltage is stepped to $-100$ V
rather than to $+100$ V, only the polarization current is observed;
there is no evidence for the injection of positive charge into the
CAS.  In addition, if negative charge has previously been injected by
applying a positive gate voltage, application of a negative gate
voltage removes little of this charge.  At $T=78$ K, the stored charge
can only be removed from the system on time scales on the order of
minutes by illumination with band-gap light.  Warming the sample to
room temperature also releases the stored charge and returns the CAS
to its initial state, although the process is considerably more
time-consuming.

The release of the stored charge when the sample is illuminated with
band-gap light suggests that the injected charge resides predominantly
on the nanocrystals, rather than within the organic capping molecules
or at defects on the (substrate) oxide surface.  This is more strongly
supported by measurements of changes in the absorption spectra with
applied voltage in sandwich structures by Woo \etal\ \cite{wing}.  For
the same electrode material and applied fields similar to those used
in this work, Woo \etal\ also observe a quenching of the nanocrystal
photoluminescence consistent with charging of the nanocrystals.

From the above data we can already eliminate two simple explanations.
First, we are measuring properties of the CAS, rather than leakage
through the substrate or conduction through the organic which fills
the interstices of the nanocrystal film.  Similar measurements on a
bare substrate show no charge injection or power-law current
transients.  Also, in several instances we have made electrical
measurements on films for which optical spectra, taken later,
showed the nanocrystals to be somewhat oxidized; these oxidized
samples, while exhibiting very weak photocurrent, showed no measurable
current injection in the dark, although the organic molecules were the
same as in the other samples.  Finally, as discussed towards the end
of this section, we see variations in the current transients with
changes in the parameters of the CAS; in future work we hope to look
at these variations in more detail.  However, we note that although
the injected charge appears to reside on the nanocrystals, at this
time we are unable to distinguish between charge which resides in the
nanocrystal core and charge which resides in a surface state of the
nanocrystal.

Second, the samples appear to be chemically stable under these
measurement conditions; several samples have been measured for over
three weeks, with no evidence for degradation.  Although there are
long-lived history effects associated with the charging of the
samples, it is possible, either by briefly warming to room temperature
or illuminating with band gap light, to remove the stored charge.  As
long as the sample is kept in vacuum and cold, the measurements for a
particular sample are reasonably reproducible.  For the remainder of the
results section we examine the details of this transient behavior as a
function of sample length, gate oxide thickness, temperature, and array 
parameters.

\begin{figure}
\epsfig{file=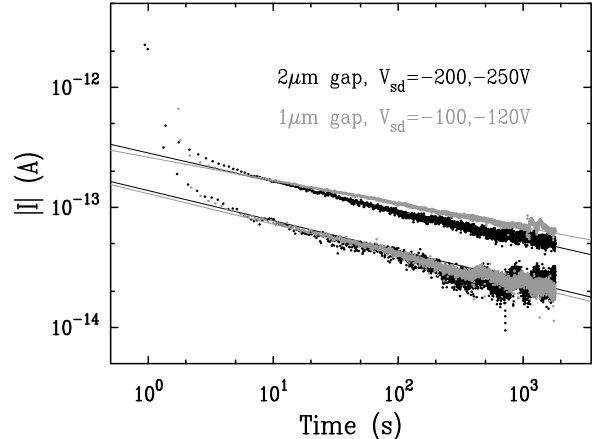, width=0.43\textwidth}
\caption{Two sets of transients for a single film of 49 \AA\
core-diameter, TOPO-capped nanocrystals, measured on a quartz
substrate with two different electrode spacings but similar applied
fields.  For the 2 $\rmu$m sample length, the transients, plotted in
black, are for steps of $V_{sd}=-200$ V and $V_{sd}=-250$ V; for the 1
$\rmu$m sample length, the transients are plotted in gray, and
correspond to steps of $V_{sd}=-100$ V and $V_{sd}=-120$ V. The solid
lines are the associated power-law fits, the parameters of which are
plotted in Fig.\ \ref{figqtzpars}.  $T=79$ K.}
\label{figqtzfdep}
\end{figure}

\begin{figure}
\epsfig{file=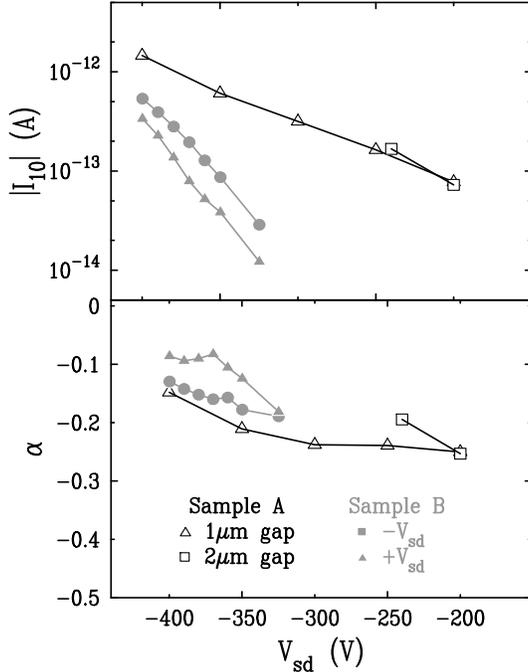, width=0.39\textwidth}
\caption{Parameters from power-law fits,  $I(t) = I_{10}(t/10s)^{\alpha}$  to current transients measured on
TOPO-capped CdSe nanocrystals deposited on a quartz substrate, plotted
as a function of applied voltage.  Top panel: the amplitude of the
current transients at ten seconds.  The approximately exponential
dependence of the transients on applied voltage is typical of
measurements made on quartz substrates.  Lower panel: the decay
exponent of the transients.  The fluctuations in the decay exponent
above $|V_{sd}|=350$ V within a single data set are within the fitting
error. Note that while the exponents are quite consistent, the
magnitude of the current and its voltage dependence are not
reproducible.  The points corresponding to the measurements of Figure
\ref{figqtzfdep} are shown in open symbols; the triangles correspond
to data for the 2 $\rmu$m sample length, and the squares to data for
the 1 $\rmu$m sample length, after the voltage has been rescaled by a
factor of two.  The agreement between the 2 $\rmu$m data and the
scaled 1 $\rmu$m data reflect the field-scaling discussed in the text.
Data for another sample, of the same diameter (49 \AA), are shown in
filled symbols, for both signs of the voltage.  Filled circles
correspond to data for $-V_{sd}$, and filled triangles to data for
$+V_{sd}$.  The discrepancy between the two signs of voltage most
likely arises from sample history effects.}
\label{figqtzpars}
\end{figure}

Just as the film accumulates charge when voltage is applied to the
gate, the film must accumulate charge at the oxide interface, possibly
across the length of the sample, when voltage is applied to the source
with the gate grounded.  Immediately after the application of a
voltage step at the source, before any charge is injected into the
film, the field is not uniform across a gated sample.  Rather, for a
particular applied voltage, the thinner the gate oxide, the stronger
the field is near the source contact.  To determine the effect of the
gate we have made measurements on gated samples with several oxide
thicknesses, as well as on quartz substrates, for which the gate is
0.5 mm distant and therefore does not affect the fields across the 1
and 2 $\rmu$m gaps.  However, we note that even for the silicon
substrates with a 350 nm gate oxide, the distance to the gate is a
substantial fraction of the electrode separation; relatively thick
gate oxides were necessary to permit the application of the large
fields necessary to pass charge through the samples.

Because there is no gate to influence the field across the electrode
spacing for the quartz substrates, it is straightforward to study the
dependence of the power-law transients on sample length.  Two pairs of
transients are plotted in Figure \ref{figqtzfdep} for a CAS on quartz;
transients measured on a sample 2 $\rmu$m long, with $V_{sd}=-200$ and
$-250$ V, are shown in black, and transients measured for a 1 $\rmu$m
long sample, with $V_{sd}=-100$ and $-120$ V are shown in gray points.
The solid lines are the corresponding fits to a power-law form,
$I(t)=I_1t^{\alpha}$.  There is reasonably good agreement between the
transients measured at the same fields for the two different sample
lengths.

The parameters of these power law fits have been plotted in Figure
\ref{figqtzpars}. The data for the 2 $\rmu$m sample length are shown
in open triangles, and the data for the 1 $\rmu$m sample, with the
voltage rescaled by a factor of two, are shown in open squares.  The
close agreement between the rescaled data for the 1 $\rmu$m sample and
that for the 2 $\rmu$m sample provide further evidence that the
current transients depend only on the applied field.

The results on quartz are generally consistent with the measurements
on the gated silicon substrates, in that we still observe power-law
current transients, and again see no evidence of steady-state current.
The major quantitative difference is that the values for $\alpha$ are
smaller than those observed for the gated substrates.  Power-law fits
to two sets of current transients measured on a quartz substrate yield
the data in Figure \ref{figqtzpars}, which shows the decay exponent,
$\alpha$, as well as the logarithm of the current at 10 s, both as
functions of the source-drain voltage.  Note that one can apply much
larger voltages than for the gated devices, because the insulator is
much thicker.  Furthermore, because the distance to the gate is much
greater than the distance between the electrodes, one expects
approximate symmetry for positive and negative voltages for the quartz
substrates, whereas we observe current transients only for negative
source voltages for the Si substrates.  The small asymmetry seen in
Fig.\ \ref{figqtzpars} for measurements on a single sample are likely
the result of sample history; we believe that some charge is stored in
the CAS even when the gate is remote.  The discrepancy between the two
different samples is somewhat more difficult to understand, but may
arise from differences in sample preparation or film thickness.

\begin{figure}
\epsfig{file=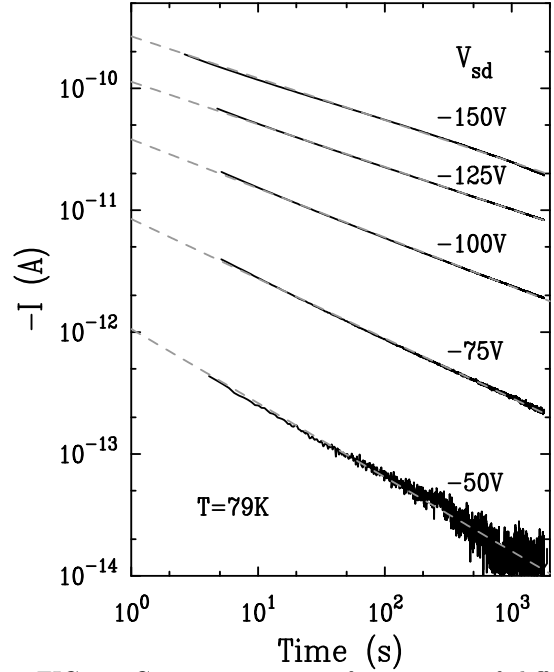, width=0.4\textwidth}
\caption{Current transients for a range of different voltage steps,
for a TOPO-capped, 45 \AA\ core-diameter sample measured on a 1
$\rmu$m gap, on a substrate with a 350 nm oxide.  The sample had been
cooled with $V_g=-150$ V several hours earlier; measurements on
samples cooled with $V_g=0$ V show similar behavior, but smaller
amplitudes for the current transients, as discussed in connection with
Fig.\ \ref{figfceff}.  Steps are made in order of increasingly
negative voltage, with a half-hour $V_{sd}=0$V step in between, during
which the sample is illuminated for twenty minutes.  Transients are
not observable for positive $V_{sd}$. $T=79$K and $V_g=0$V
throughout. Dashed lines are power-law fits to the data; the
parameters from these fits are plotted in Fig.\ \ref{figIVpars}.}
\label{figIV1}
\end{figure}

\begin{figure}
\epsfig{file=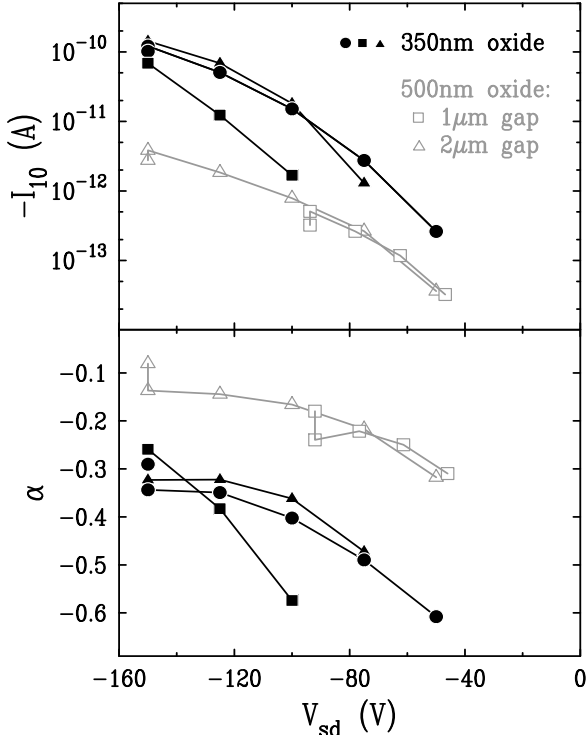, width=0.43\textwidth}
\caption{Parameters from the power-law fits $I=I_1t^{\alpha}$ to
current transients measured for samples on gated substrates.  Top
panel: the amplitude of the transients at ten seconds.  Lower panel:
the decay exponent of the transients.  Data for 1 $\rmu$m long samples
on substrates with a 350 nm gate oxide are shown in black; the filled
circles correspond to the dashed lines in Fig.\ \ref{figIV1}, and the
triangles come from measurements taken on a different but nominally
identical sample. The filled squares correspond to measurements on the
first sample, immediately after cooling with $V_g=0$V.  Data for a
sample of 45 \AA\ diameter nanocrystals measured on a substrate with a
500 nm oxide substrate are shown in gray, for two different sample
lengths.  The open triangles correspond to the 1 $\rmu$m sample, and
the open squares are data for the 2 $\rmu$m long sample, after
dividing the voltage by a factor of 1.6.  Transients are only observed
for negative $V_{sd}$ on the gated substrates. }
\label{figIVpars}
\end{figure}

As seen in Figure \ref{figqtzpars}, the current at fixed time
increases exponentially with the applied voltage for the CAS on a
quartz substrate. For the gated samples, however, the voltage
dependence is more complex.  Figure \ref{figIV1} shows transients for
a range of voltages, measured for a sample on a 350 nm oxide.  For
similar voltages, the currents at 10 s are considerably larger for the
gated substrates, allowing us to measure transients at lower voltages;
this is probably the result of higher fields in the sample near the
source electrodedue to the presence of the gate.  In Fig.\
\ref{figIVpars} the parameters obtained from fits of a power-law decay
to these transients are plotted, together with a set of points
corresponding to a nominally identical sample with a similar history,
all in filled black symbols.  There is remarkably good agreement
between the two sets of data.  On the same axes are also plotted
another set of points, which correspond to a set of transients
measured for a sample on a substrate with a 500 nm oxide.  For both
types of gated substrate, the decay exponents become smaller with more
negative $V_{sd}$, and the amplitude of the current at 10 s, as well
as the exponent $\alpha$, appear to saturate as the size of the
voltage step is increased.

Although the current transients scale with field on the quartz
substrates, the scaling with applied voltage is different for samples
measured on gated substrates \cite{fn4}.  For a 500 nm thick oxide, although we
can find a voltage such that the transient on the 2 micron gap is the
same as that on the 1 micron gap, the voltages differ by a factor of
1.6 instead of 2. Interestingly, this single rescaling factor
persists over the entire voltage range for which the transients are
measurable.  This is illustrated in Fig.\ \ref{figIVpars}, where the
data for the 500 nm oxide substrates consist of data for a 1$\rmu$m
sample, shown in open triangles, as well as rescaled data for a 2
$\rmu$m sample, shown in open squares. The scaling factor appears to
be smaller still for the 350 nm oxide.  Thus it appears that the
transients on quartz substrates depend only on the field, whereas for
the silicon substrates, as the gate oxide thickness is reduced, the
transients become less dependent on the electrode spacing, perhaps
approaching a dependence only on the source voltage.

\begin{figure}
\epsfig{file=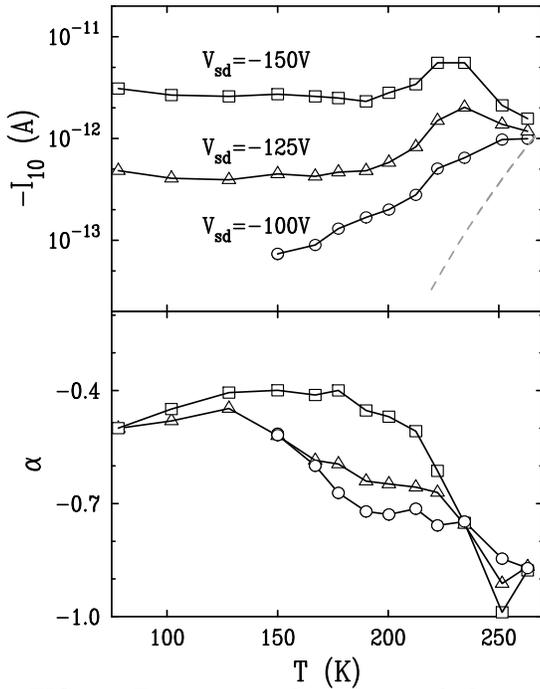, width=0.4\textwidth}
\caption{Temperature dependence of the current transients measured on
the same sample as in Fig.\ \ref{figIV1}, taken in order of increasing
temperature.  The two parameters are obtained from three-parameter
fits to the transients, $I(t)=I_{10}(t/10s)^{\alpha}+I_{\mathit
leakage}$, where $I_{\mathit leakage}=0$ for $T<200$ K.  Top panel:
the amplitude of the current transients at ten seconds.  The dashed
line indicates the value of $I_{ss}$ for $V_{sd}=-100$ V, which is
identical to that measured on a bare substrate, and so is attributed
to oxide leakage.  Lower panel: the decay exponent of the transients,
for the same values of $V_{sd}$. Note that the transients for
$V_{sd}=-100$ V are too small to fit below 150 K.}
\label{figtempdep}
\end{figure}

The temperature dependence of the current transients is shown in
Figure \ref{figtempdep}, for measurements of a 1 $\rmu$m long sample
on a substrate with a 350 nm gate oxide; these parameters are obtained
from three-parameter power-law fits to the transients,
$I(t)=I_1t^{\alpha}+I_{\mathit leakage}$.  In the top panel, the
current at ten seconds is plotted for three different values of
$V_{sd}$.  Also shown is the steady-state leakage current, which is
identical to that measured on an equivalent device structure with no
CAS film.  This leakage current is linear in voltage, and is well
described by an activated form, $I=I_0\exp(-E_A/k_BT)$, with an
activation energy $E_A=370$ meV.  Clearly, the leakage current
interferes with the measurements of the transients at temperatures
above $\sim220$ K.  The decay exponent of the current transients is
plotted in the lower panel.  Although Fig.\ \ref{figtempdep} shows
only data from 77K to room temperature, a separate set of measurements
from 6K to 77K shows no substantial variation with temperature over
that range.  Above 77K, there is little variation in either of the
parameters until approximately 150K, above which the amplitude of the
current transients increases and the power law becomes steeper.  The
decrease in the amplitude of the transients above 220 K, and the
associated sharp drop in the decay exponent, is most likely an
artifact associated with the leakage current.

In order to further explore the relationship between the buildup of
charge in the CAS and the current transients, we have measured
transients after cooling the samples with an applied electric field.
At room temperature, a voltage $V_{fc}$ is applied to the gate, and
the sample is then cooled to 77 K.  Once the sample is cold, the gate
voltage is turned to zero, and only then is the source voltage
stepped.  From the measurements of the temperature dependence of the
transients, it appears that charge injected into the film equilibrates
fairly rapidly at room temperature.  Furthermore, the measurements of
gate charging, shown in Fig.\ \ref{figgatecharge}, demonstrate that
once charge is injected into the CAS, it is difficult to remove at low
temperatures.  Therefore, in applying a gate voltage at room
temperature, the hope is to equilibrate the sample with a particular
charge density at room temperature, and then to maintain that charge
density after the gate voltage is turned to zero with the sample cold.

\begin{figure}
\epsfig{file=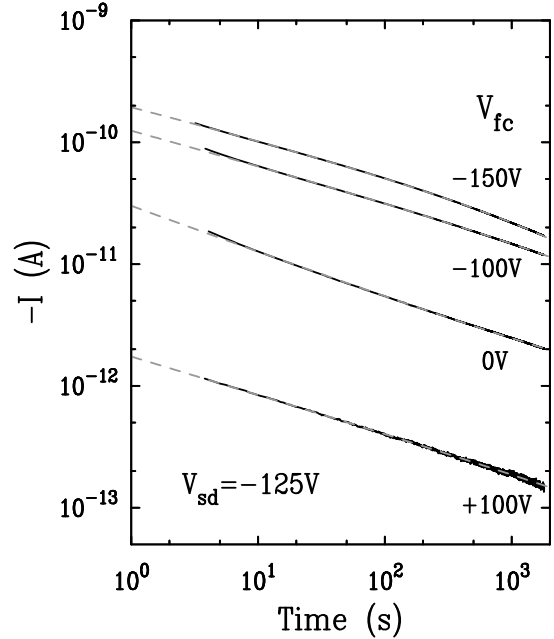, width=0.4\textwidth}
\caption{Current transients for the same sample as in Fig.\
\ref{figIV1}, after cooling with different gate voltages, shown with
power-law fits (dashed lines).  Before each transient, the sample has
been prepared by cooling from approximately 290K to 78K with $V_{fc}$
(labelled on the right) applied to the gate; after reaching 78K, $V_g$
is set to 0V, and then $V_{sd}$ is stepped to $-125$ V.}
\label{figfceff}
\end{figure}

The results, for a set of transients measured with $V_{sd}$ steps to
$-125$ V, are shown in Figure \ref{figfceff}.  When the sample is
cooled with a positive voltage on the gate, such that excess negative
charge has been stored in the sample, the current transient is much
smaller; conversely, cooling with a negative voltage on the gate
results in a much larger current.  This is consistent with the decay
of the current in time; in both cases, the buildup of negative charge
increases the resistance.  It also suggests that at room temperature
it may be possible to inject positive charge, although, given the
leakage current observed even for the bare substrates (Fig.\
\ref{figtempdep}), it remains possible that some of the charge
injected at room temperature resides in the oxide rather than within
the nanocrystal film.

\begin{figure}
\epsfig{file=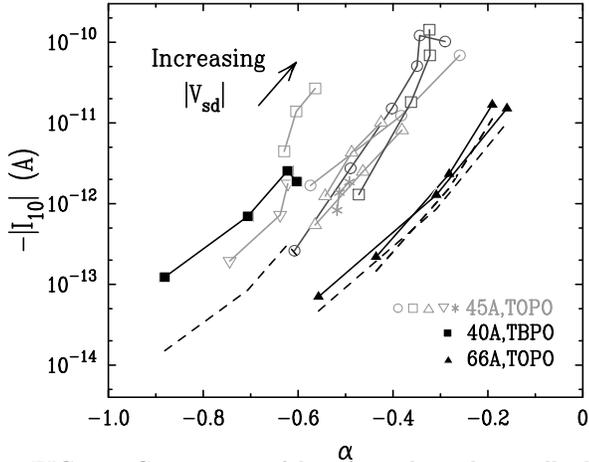, width=0.43\textwidth}
\caption{Comparison of data from three chemically distinct types of
samples measured on substrates with 350 nm oxides, in a plot of the
amplitude of the transients at ten seconds against the exponent of the
decay, obtained from power-law fits.  In gray symbols are data from
five different 45 \AA\ core diameter, TOPO-capped samples, after
scaling for the film thickness as described in the text.  Data for the
other two samples are shown before rescaling as dashed lines, and
after rescaling, as filled black squares for the 40 \AA\ core-diameter
TBPO-capped sample, and filled black triangles for a 66 \AA,
TOPO-capped sample.  The trends discussed previously, with applied
voltage and temperature, are indicated with arrows.}
\label{figarrpar}
\end{figure}

The parameters of the current transients depend on the size and
spacing of the nanocrystals in the CAS.  In Figure \ref{figarrpar} we
look for trends with the sample parameters by plotting the amplitude
of the current transient against the decay exponent $\alpha$, for a
number of different samples with 1 $\rmu$m length measured on
substrates with 350 nm oxides.  Each line connects a series of
measurements on a single sample with different applied voltages
($V_{sd}$).  The general direction of increasingly negative $V_{sd}$
is indicated by the arrows.  Data for the most commonly used type of
sample, 45\AA\ core-diameter TOPO-capped nanocrystals, are shown for
five different samples.  Although there is some spread among these
points, data for the other two types of samples fall outside this
spread.  Data for a sample of 66\AA\ core-diameter, TOPO-capped
nanocrystals show somewhat smaller values for $|\alpha|$ as well as
the amplitude of the transients, corresponding to a slower decay of
the current.  Data for a sample of 40\AA\ core-diameter, TBPO-capped
nanocrystals show larger values of $|\alpha|$, as well as smaller
amplitudes.

In making comparisons between different samples, it is necessary to
take the varying thicknesses of the films into account; the data in
Fig.\ \ref{figarrpar} correspond to films which range in thickness
from 100 to 900 nm, corresponding to 20 to 180 layers of nanocrystals.
As a first approximation, we have assumed that the current is
proportional to the thickness of the film.  The data for the standard
samples in Fig.\ \ref{figarrpar} has already been scaled to correct
for the thickness; the clustering of these points indicates that the
simple multiplicative correction is moderately successful, and
suggests that the current distribution is roughly uniform for these
films.  There is, nonetheless, one major difference for the thinner
films, shown as triangles and asterisks; although the scaled points
fall in the same general area as those of the thicker samples, they
span a much smaller range of $\alpha$ and $I_{10}$ for the same range
in $V_{sd}$, suggesting a change in the voltage dependence of the
transients.  The dependence of the current transients on film
thickness requires future attention, and could be particularly
interesting in the event that the two-dimensional limit could be more
closely approached.

\subsection*{Discussion}

The very high resistivity of our CAS films suggests that they contain
no free carriers in the initial state.  Since such free carriers are
thought to quench photoluminescence \cite{chepic}, the high
resistivity is therefore consistent with the relatively high quantum
efficiencies found in similar films.  However, our experiments show
that at sufficiently high voltages negative charge can be injected
into the CAS from gold electrodes.  Furthermore, once the charge is
injected, it remains in the film for very long times at temperatures
below 200K.  The essential questions raised by our experiments are why
the charge is stored for such long times and what the relationship is
between this stored charge and the observed power-law decay of the
current.

We first focus on the quartz-substrate measurements, for which the
gate affects neither the field nor the charge accumulation in the
sample.  The exponential voltage dependence of the current at fixed
time might result from field emission of electrons from the Au into
the CAS.  In this case, one might argue that a buildup of charge near
the source electrode would screen the field, reducing the the field
emission.  Ginger \etal\ have developed such a model to explain their
observed current transients for CdSe nanocrystal arrays \cite{ginger}.
Using a sandwich-cell geometry, with sample thicknesses of 100-200 nm,
they find a current that decays as a stretched exponential in time.
In this model, charge carriers move in a transport band until they
become trapped in localized states.  The buildup of charge reduces the
field at the contact and consequently the injection current.

We find it difficult to reconcile this model with our observations.
First, it is difficult to understand why extrinsic traps should lead
to current transients whose exponents change with the nanocrystal size
and spacing.  Second, in our measurements, orders of magnitude more
charge flows to the drain than is trapped in the sample. Since the
trapping probability must therefore be small, the density of trapped
charge should be approximately uniform across the sample.  At later
times, when this uniform density of trapped charge becomes sufficient
to prohibit injection from the contact, the current would not scale
with the applied field.  However, the current transients we observe
are field-dependent at all times; there is no evidence for a
significant buildup of charge within the sample during the
measurement.  Furthermore, because the amplitude of the transients
grows exponentially with the applied field, these measurements are
fairly sensitive to deviations from field-dependence.

Alternatively, the exponential voltage dependence could arise from an
exponential dependence of the hopping rate between nearest-neighbor
nanocrystals.  In this case, we propose that the current is not
limited by injection into the sample, but rather by extraction from a
space charge region which forms near the contact.  This type of
space-charge limited current is common in insulators with low charge
density; however, for the CAS our observations differ from the
conventional models of space charge limited current in two ways.
First, in most systems, space-charge limited currents reach steady
state when the integrated current is roughly equal to the total amount
of space charge which resides in the sample.  For the CAS on quartz
substrates, we estimate this time to be $t<1$ s, much shorter than the
times over which we observe the power law decay.  From this, we are
led to propose that although the space charge region develops fairly
quickly, the extraction of carriers from the space charge region
decreases with time. 

Second, for conventional space charge limited current, although the
space charge is concentrated near the injecting contact, there is
typically a long tail in the distribution which extends across the
sample.  As a result, in systems where space charge limited current is
observed, the current generally has some dependence on the sample
length as well as on the applied field \cite{kao.hwang}.  However, in
the CAS films we observe field-dependence of the current transients,
over a time range of 1000 s for measurements on the quartz substrates.
This suggests that the space charge region which limits the transport
in this system is confined to a narrow region near the injecting
contact.  From the uncertainty associated with the scaling factor in
our measurements, we estimate that this region extends no further than
100 nm from the contact, or a tenth of our smallest sample length.  We
note that our observations are not entirely inconsistent with those of
Ginger \etal , whose samples were considerably shorter than those
studied here; a space charge region of 100 nm in their system would
extend halfway across the longest samples.

The power-law decay of the current is reminiscent of that associated
with relaxation of carriers in GaAs \cite{monroe}.  The motion of
electrons in a band of localized states with disorder has been studied
in great detail \cite{efros.shklov}. This system has a Coulomb gap for
single-particle excitations, and the ground state is highly
degenerate.  Together, these give rise to very slow relaxation times,
and the system has been referred to as a Coulomb glass \cite{dlr}.
Such slow relaxation has been suggested as the origin of unusual
phenomena in indium oxide films \cite{ovad.prb,ovad.prl}.  We propose
that the space charge near the source electrode is a Coulomb glass.
Although the space charge builds up in a very short time, the emission
of electrons from the space charge region decreases with time as the
Coulomb glass relaxes and the Coulomb gap grows in size.  Presumably,
once such a Coulomb glass relaxes it will be removed from the sample
only extremely slowly after the voltage is removed; this is consistent with
the sample history effects we observe as well as the charging
measurements of Fig.\ \ref{figgatecharge}.  The larger exponents
observed as the temperature is increased might result from more rapid
relaxation of the Coulomb glass.  A similar trend towards faster decay
of the current is seen when changes in the array parameters lead to
increased coupling between the dots, in Fig.\ \ref{figarrpar}.  For
the TBPO-capped sample, the nearest-neighbor separation is 7\AA ,
compared to 11\AA\ for the standard, TOPO-capped dots \cite{murray.sci}.
Because inter-dot tunneling should increase exponentially as the width
of the tunnel barrier is decreased, we expect the change in capping
molecules to lead to increased coupling between the dots.  As for the
temperature dependence, the data suggest that increased coupling
between the dots leads to a faster decay of the current.

We have no quantitative theory with which to compare our results, and
there are even qualitative observations that we cannot explain.  For
example, it is clear that the presence of a gate electrode will lead
to additional space charge in the sample, so it is not surprising that
the current transients are strongly modified as the gate oxide is made
thinner than the spacing between source and drain.  The field which
extracts charge from the space charge region near the injecting
contact then falls between source and gate; it increases as the gate
oxide thickness decreases, and becomes less dependent on the electrode
spacing.  This is consistent with the data of Fig.\ \ref{figIVpars},
for which $I(t, V, L)=I(t, 1.6V, 2L)$ for the 1 $\rmu$m and 2 $\rmu$m
long samples on the gated substrates, rather than the strict field
dependence observed for the quartz substrates.  However, we do not
understand why the exponents of the current decay are larger for the
gated substrates; in the Coulomb glass model the presence of more
space charge should lead to slower equilibration. We also do not know
how to reconcile the smaller exponents of the power-law decay of the
current when we apply the voltage to the source compared with that
which we observe when voltage is applied to the gate.

Our future work will focus on making samples with stronger coupling between
nanocrystals.  Presumbably this would make it possible to observe the
hopping conductivity characteristic of the Coulomb gap.

\begin{acknowledgments}
This work at MIT was supported by Award No. DMR 9808941, under the
MRSEC Program of the NSF.  N.Y.M. acknowledges support from an ONR
Graduate Fellowship and from the Lucent Technologies Graduate Research
Program for Women.
\end{acknowledgments}

\bibliographystyle{prsty}

\begin{thebibliography}{10}
\bibitem{fn0}
Other measurements, made on substrates with circular electrodes, are
consistent with those shown here; for these devices, the active area
of the film lies inside the outer electrode, and so the similarity of
the results confirms this picture.
\bibitem{fn1} 
A stretched exponential decay will always be concave downward on a
log-log plot, although with $\tau,\beta \rightarrow 0$ the form
approaches a straight line.  Here we choose a characteristic time of
$\tau = 1$ ms, an order of magnitude shorter than our earliest
measurement time; longer values of $\tau$ would show still more
curvature.  In the data from the sandwich structures of Ref.\
\cite{ginger}, typical values of $\tau$ were 10-500 seconds.
\bibitem{fn2} 
Although charge spreads into the film in all directions from the
source electrode, the measured current arises only from charge which
either passes into, or moves near to (relative to the thickness of the
gate oxide), the drain electrode.  More distant charge will be
screened by the gate electrode, which is grounded for the measurement.
Therefore, when estimating the possible contribution of stored charge
to the measured current, we consider only the area of the film between
the electrodes.  Even if the entire surface of the substrate, except
for the drain electrode, were charged to $V_{sd}$, this measurement
would be sensitive only to charge around the perimeter of the
electrode, leading to a correction of less than a factor of ten.
\bibitem{fn3}
In photoconductivity measurements on similar samples \cite{calphoto},
photocurrent several orders of magnitude larger than the dark current
is observed.  The spectral dependence of the photocurrent is the same
as the optical absorption of the nanocrystals.  In the measurements
discussed here, the creation of electron-hole pairs in the
nanocrystals apparently makes the CAS sufficiently conducting for the
charge to return to ground.
\bibitem{fn4} 
Here, we use the word scaling in the general sense: to say that the
transients scale with the field is to say that $I(t)$ depends only on
$E=V_{sd}/L$, where $L$ is the sample length, rather than $I(t)\propto E$.
\end{thebibliography}

\newcommand{\noopsort}[1]{} \newcommand{\printfirst}[2]{#1}
  \newcommand{\singleletter}[1]{#1} \newcommand{\switchargs}[2]{#2#1}

\end{document}